\documentstyle[aps,epsf]{revtex}

\newcommand{\calu}{{\cal U}}
\newcommand{\calw}{{\cal W}}

\newcommand{\ie}{i\epsilon}
\newcommand{\np}{\not\! p}
\newcommand{\lan}{\langle }
\newcommand{\ran}{\rangle }

\def \p{_{\perp }}

\newcommand{\iy}{\infty }
\newcommand{\beq}{\begin{equation}}
\newcommand{\eeq}{\end{equation}}
\newcommand{\defi}{\stackrel{\rm def}\equiv}
\newcommand{\bega}{\begin{eqnarray}}
\newcommand{\ega}{\end{eqnarray}}
\newcommand{\Repa}{\mathop{\rm Re}}         % Real Part
\newcommand{\Tr}{\mathop{\rm Tr}}                %Trace
\newcommand{\al}{\alpha }

\newcommand{\Ud}{U^{\dagger }}
\newcommand{\Vd}{V^{\dagger }}
\newcommand{\Wd}{W^{\dagger }}

\newcommand{\si}{\sigma }

\newcommand{\half}{\frac {1}{2}}

\begin{document}

\begin{flushright}
  JLAB-THY-97-22\\
  19 June 1997 
\end{flushright}
\vspace{1 cm}
\begin{center}
{\Large \bf Operator Expansion For Diffractive
High-Energy Scattering
\footnote{Talk presented at 
5th International Workshop on  Deep Inelastic Scattering
and QCD (Chicago, April 1997)}}
\end{center}
\begin{center}
{I.I.  BALITSKY\footnote{Also at St. 
Petersburg Institute of Nuclear Physics,
Gatchina, Russia}} \\
{\em Physics Department, Old Dominion University,}
\\{\em Norfolk, VA 23529, USA}
 \\ {\em and} \\
{\em Jefferson Lab,} \\
 {\em Newport News,VA 23606, USA}
\end{center}

\begin{abstract}

 I discuss the operator expansion for diffractive 
high-energy scattering and present the non-linear 
evolution equation for the relevant Wilson-line operators 
which describes both the propagator of the BFKL pomeron and the 
three-pomeron vertex.
\end{abstract}
\vspace{1cm}
Semihard diffractive processes are
of special interest since they provide us with valuable information
about the non-linear dynamics of the pomerons. 
(For review of the experimental situation, see ref. \cite{difhera}).
In a recent paper\cite{ing} I suggested the operator expansion (OPE) 
for high-energy amplitudes. It turns out that the small-$x$ behavior 
of structure functions of DIS
is governed by the evolution of Wilson-line operators with respect
to the deviation of the supporting lines from the light cone.
In this paper, I generalize this approach to diffractive 
high-energy scattering. In particular, I obtain the cross section of
the diffractive dissociation of the virtual photon in the 
triple Regge limit $s\gg M^2\gg Q^2$ as a result of the evolution of the
relevant Wilson-line operators.

\section*{OPE for high-energy amplitudes}

First, let me remind the OPE for high-energy amplitudes derived 
in  \cite{ing}. Consider the amplitude of forward 
$\gamma^*\gamma^*$-scattering at small $x_B={Q^2\over s}$.
In the target frame, the virtual photon splits into $q\bar{q}$ pair
which approaches the nucleon at high speed. Due to the high 
speed  the classical trajectories of the quarks are straight lines
collinear to the momentum of the incoming photon $q$. 
The corresponding operator expansion 
switched between nucleon states has the form \cite {ing}:
\begin{eqnarray}
&\int \! d^{4}x
    e^{iq\cdot x} \lan p|T\{j_{\mu }(x)j_{\nu }(0)\}|p\ran
= \int d^{2}x\p
   I_{\mu\nu}(x\p)\lan p|\Tr\{\hat{U}(x\p)\hat{\Ud}(0)\} |p\ran,
\label{1.1}
\end{eqnarray}
where $I_{\mu\nu}(x\p)$ is a certain numerical 
function of the transverse separation of quarks $x\p$ and virtuality
of the photon $Q^2=-q^2$. The relevant
operators $U(\Ud)$ are gauge factors ordered along the classical
trajectories which are almost light-like lines stretching from minus to 
plus infinity:
\begin{equation}
U(z\p )= P\exp\left(i\int_{-\iy}^{\iy}due^{\mu}A_{\mu}(ue+z\p)\right)
 \label{1.2}
 \end{equation}
where $e$ is collinear to $q$ and $z\p$ is the transverse position
of the Wilson line. 

It
turns out that the small-$x$ behavior of structure functions
is governed by the evolution of these operators with respect
to the deviation of the Wilson lines from the light cone; this
deviation serves as a kind of ``renormalization point" for these operators.
At infinite energy, the vector $e$ is light-like and
the corresponding matrix elements of the operators (\ref{1.2}) have a
logarithmic divergence in longitudinal momenta. To regularize it , 
we consider
operators corresponding to large but finite velocity and take 
$e_{\zeta}=e_1~+~\zeta e_2$
where $e_1=(q-\frac{q^2}{2pq}p)$ and $e_2=p$ are the lightlike 
vectors close to the directions of the colliding particles. 
Now, instead of studying the energy-dependence of the 
physical amplitude we must investigate the dependence of the operators
(\ref{1.2}) on the slope $\zeta$. Large energies mean small 
$\zeta$ and we can
sum up logarithms of $\zeta$ instead of logarithms of $s$
(At present, we can do it only in the leading logarithmic 
approximation (LLA) $\al_s\ll 1$, $\al_s\ln{s\over m^2}\sim 1$).
The equation governing
the dependense of $U$ on $\zeta$ has the form \cite{ing}:
\begin{eqnarray}
\lefteqn{\zeta \frac{d}{d\zeta} \calu(x\p,y\p)={3\al_s\over 2\pi^2} \int dz\p
{(x\p -y\p )^2\over (x\p -z\p )^{2}(z\p -y\p )^2}}\nonumber\\
&\left\{\calu(x\p,z\p)
+\calu(z\p,y\p)-\calu(x\p,y\p)+\calu(x\p,z\p)\calu(z\p,y\p)\right\}
\label{1.5}
\end{eqnarray}
where $\calu(x\p,y\p)\defi{1\over 3}\Tr\{U(x\p )\Ud (y\p )\}-1$.
The first three linear terms in braces in the r.h.s. of eq. (\ref{1.5}))
reproduces the BFKL pomeron\cite{bfkl} while the quadratic term will give us the 
three-pomeron vertex as we shall see below. 
The solution of the linearized evolution equation is especially simple in 
the case of zero momentum transfer  (e.g. for the total 
cross section of small-x DIS):
\bega
&\lan p|\calu^{\zeta=x_B}(x\p,0)|p\ran=
\int \! \frac {d\nu }{2\pi ^{2}}
(x\p^{2})^{-\half+i\nu }
\left(\frac {s}{m^{2}}\right)^{\omega(\nu )}
\int \! dz\p(z\p^{2})^{-\half-i\nu }
\lan p|\calu^{\zeta_0}(z\p,0)|p\ran
\label{1.10}
\ega 
where 
$\omega(\nu )=6{\alpha _{s}\over\pi }[-\Repa\psi (\half+i\nu )-C]$
and
$m^2$ is either $Q^2$ or $m_N^2$ (in LLA, we cannot distinquish
between $\al_s\ln{s\over Q^2}$ and $\al_s\ln{s\over m_N^2}$).
The sketch of linear evolution is presented in Fig. 1
%======================================================
\begin{figure}[htb]
\vspace{0cm}
\mbox{
\epsfxsize=13.5cm
\epsfysize=4cm
\hspace{0cm}
\epsffile{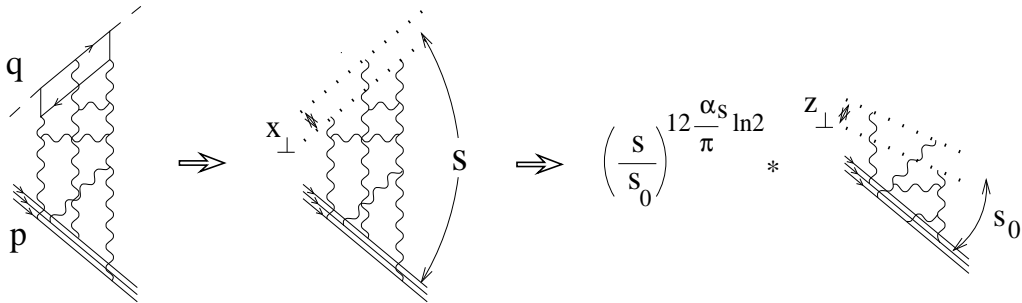}}
\vspace{0.5cm}
{\caption{\label{fig:1}BFKL evolution in terms of Wilson-line operators
(denoted by dotted lines).}}
\end{figure}
The starting point of the evolution is the slope collinear 
to the momentum of the incoming photon q ($\zeta=x_B$)
and it is convenient to stop the evolution at a certain intermediate 
point $\zeta_0={Q^2\over s_0}$ where 
$s_0\gg m_N^2, ~~{\al_s\over\pi}\ln{s_0\over m_N^2}\ll 1~$.
The first of these conditions means that $s_0$ is still high from the 
viewpoint of low-energy
nucleon physics while the second condition means that $s_0$ is sufficiently
small from the viewpoint of high-energy physics 
(so one can neglect BFKL logs).
The matrix element of the double-Wilson-line operator at this
slope is a phenomenological input for the BFKL evolution (just as the 
structure function at low $Q^2$ serves as the input for ordinary DGLAP evolution).
At large $s$ the integral over $\nu$ is dominated by the vicinity of $\nu=0$
which gives the familiar BFKL asymptotics:
\beq
\si^{\rm tot}\simeq x_B^{12{\al_s\over \pi}\ln 2}
\label{1.11}
\eeq

Note, however, that the full nonlinear equation (\ref{1.5})
contains more information than the linear BFKL equation --- for example,
it describes also the triple vertex of hard pomerons in QCD.
In order to see that, it is convenient to consider some process
which is dominated by the three-pomeron vertex  --- the best
example is the diffractive dissociation of the virtual photon.

\newpage
\section*{OPE for diffractive cross sections}
The total cross section for diffractive scattering
has the form:
\beq
\si^{\rm diff}_{\rm tot}=\int dx e^{iqx}\int {d^3p'\over(2\pi)^3}\sum_X
\lan p|j_{\mu}(x)|p'+X\ran \lan p'+X|j_{\nu}(0)|p\ran
\label{1.12}
\eeq
where $\sum_X$ means the summation over all intermediate states.
We can formally write down this cross section 
as a ``diffractive matrix element''(cf ref. \cite{bbi}):
\bega
&\si^{\rm diff}_{\mu\nu}=\lan p|T\{j^-_{\mu}(x)j^+_{\nu}(0)
e^{i\int dz({\cal L}^+(z)-{\cal L}^-(z))}\}|p\ran_{\rm diff}
\label{1.14}
\ega
The index ``$-$'' marks the fields to the left of the cut and ``$+$'' to the
right. The definition of the T-product of the 
fields with $\pm$ labels is as follows: the ``$+$''fields are time-ordered, 
the ``$-$''fields stand 
in inverse time order (since 
they correspond to the complex conjugate amplitude), and ``$-$''
fields stand always to the left of the ``$+$''ones. Therefore, the
diagram technique with the double set of fields is 
the following: contraction of two ``$+$''fields is the usual 
Feynman propagator ${\not p\over p^2+\ie}$ (for the quark field), 
contraction of two $``-''$ 
fields is the complex
conjugated propagator ${\not p\over p^2-\ie}$, and the contraction of 
the ``$-$'' field with the ``$+$''one is the ``cutted propagator''
$2\pi\delta(p^2)\theta(p_0)\!\np$
\footnote{
We  use the $-+$ perturbative propagator only for hard momenta 
so the additional emitted nucleon with momentum $p'$ 
(constructed from soft quarks) can be  factorized.
}. 
This diagram technique for calculating T-products of double sets of 
fields exactly reproduces the Cutkosky rules for the calculation 
of cross sections.

The main result of this paper is the operator expansion for the
diffractive amplitude $\si^{\rm diff}_{\mu\nu}$. Similarly to the case
of the usual amplitude (\ref{1.1}), we get in lowest order in $\al_s$:
\begin{eqnarray}
&\si^{\rm diff}_{\mu\nu}=
 \int d^{2}x\p I_{\mu\nu}(x\p)\lan p|\Tr\{W(x)\Wd(0)\} |p\ran_{\rm diff},
\label{1.16}
\end{eqnarray}
Here
$W(x\p)=\Vd(x\p)U(x\p)$ 
where $U(x\p)$ denotes the  Wilson-line operator (\ref{1.2})
constructed from $``+''$ fields and $V(x\p)$ from $``-''$ fields.

The evolution equation (with respect to the slope of the supporting line)
turns out to have the same form as eq. (\ref{1.5}) for usual amplitudes:
\begin{eqnarray}
\lefteqn{\zeta \frac{d}{d\zeta} \calw(x\p,y\p)={3 \al_s\over 2\pi^2}\int dz\p
{(x\p -y\p )^2\over (x\p -z\p )^{2}(z\p -y\p )^2}}\nonumber\\
& \left\{\calw(x\p,z\p)
+\calw(z\p,y\p)-\calw(x\p,y\p)+\calw(x\p,z\p)\calw(z\p,y\p)\right\}
\label{1.19}
\end{eqnarray}
where $\calw(x\p,y\p)\defi{1\over 3}\Tr\{W(x\p )\Wd (y\p )\}-1$.
Consequently, the linear evolution has the same form as (\ref{1.10}).

Let us describe now the diffrractive amplitude in 
LLA and in leading order in $N_c$. In this approximation 
we must take into account the non-linearity in eq. (\ref{1.19}) 
only once, and the rest of the evolution is linear. The result is
(roughly speaking) the three two-gluon BFKL ladders 
which couple in a certain point --- see Fig.2.
%======================================================
\begin{figure}[htb]
\vspace{0cm}
\mbox{
\epsfxsize=6cm
\epsfysize=6cm
\hspace{4cm}
\epsffile{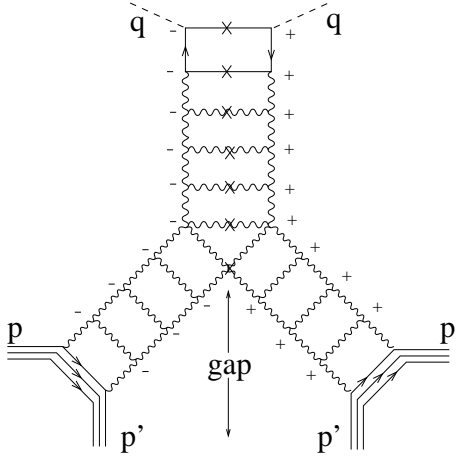}}
\vspace{0.5cm}
{\caption{\label{fig:fig2}Amplitude of diffractive scattering in the LLA-$N_c$
approximation.}}
\end{figure}
For the case of diffractive DIS, 
this evolution has the form (cf. ref. \cite{bartels}):
\bega
\lefteqn{\lan p|\int dy\p \calw^{\zeta=x_B}(x\p+y\p,y\p)|p\ran=}\nonumber\\
&\displaystyle{3\al_s \over 8\pi^3}\int d\nu (x\p^2)^{\half+i\nu}
\int d\nu_1 d\nu_2 dx_{1\perp} dx_{2\perp}\nu_1^2\nu_2^2 
\left((x_1-x_2)\p^2\right)^{-\half+i(\nu_1+\nu_2-\nu)}
\nonumber\\
&\displaystyle\Theta(\nu;\nu_1,\nu_2)\int{d^2p'\p\over 4\pi^2} 
\int^s_{Q^2}{dM^2\over M^2}\left({s\over M^2}\right)^{\omega(\nu)}
\left({M^2\over Q^2}\right)^{\omega(\nu_1)+\omega(\nu_2)}
\nonumber\\
&\lan p|\calu^{\zeta_0}(x_{1\perp},\nu_1|p'\ran
\lan p'|\calu^{\zeta_0}(x_{2\perp},\nu_2)|p\ran
\label{1.22}
\ega
where $M^2$ is the invariant mass of the produced particles,
\bega
&\calu(x\p,\nu)= \int dx\p' dy\p'
\left({(x'-y')\p^2\over (x'-x)\p^2(y'-x)\p^2}\right)^{\half+i\nu}
{\calu(x\p',y\p')\over(x'-y')\p^4}
\label{1.23}
\ega
is the eigenfunction\cite{lip} of the linear evolution equation (\ref{1.5}) 
(at $t\not =0$)
and $\Theta(\nu;\nu_1,\nu_2)$ is a certain numerical function of
three $\nu$'s. (The coupling constant of three BFKL pomerons (\ref{1.11})
is $\Theta(0;0,0)\simeq 1.58$).
The value of $M^2$ determines the
rapidity gap: from $\eta=\ln{s\over Q^2}$ to  $\eta=\ln{M^4\over Q^2s}$ 
we have a production of particles 
described by  the cut part of the ladder in Fig. 2 
which brings in the factor $\left({s/M^2}\right)^{\omega(\nu)}$ while
from $\eta=\ln{M^4\over Q^2s}$ to $\eta=\ln x_B$ we 
have a rapidity gap so there are two 
independent BFKL ladders which bring in the factors 
$\left({M^2/Q^2}\right)^{\omega(\nu_1)}$ and 
$\left({M^2/Q^2}\right)^{\omega(\nu_2)}$.
The coupling of the BFKL pomeron with non-zero momentum transfer to 
the nucleon is decribed by the matrix element $\lan p'|\calu(x,\nu)|p\ran$.
At high energies and momentum transfer, it can be approximated by the 
non-forward gluon parton density.
\section*{Acknowledgments}
This work was supported by the US Department of Energy under
contract DE-AC05-84ER40150.

\end{document}